\documentclass[onecolumn,showpacs]{revtex4}%
\usepackage{amssymb}
\usepackage{amsmath}
\usepackage{epsfig}
\usepackage{amsfonts}
\usepackage{graphicx}%
\begin{document}
\title{Nuclear matter within a dilatation-invariant parity doublet model: the role of
the tetraquark at nonzero density }
\author{Susanna Gallas$^{\text{a}}$, Francesco Giacosa$^{\text{a}}$, and Giuseppe
Pagliara$^{\text{b}}$}
\affiliation{$^{\text{a}}$Institute for Theoretical Physics, Johann Wolfgang Goethe
University, Max-von-Laue-Str.\ 1, D--60438 Frankfurt am Main, Germany}
\affiliation{$^{\text{b}}$ Institut f\"{u}r Theoretische Physik,
Ruprecht-Karls-Universit\"at, Philosophenweg 16, D-69120, Heidelberg, Germany}

\begin{abstract}
We investigate the role of a scalar tetraquark state for the description of
nuclear matter within the parity doublet model in the mirror assignment. In
the dilatation-invariant version of the model a nucleon-nucleon interaction
term mediated by the lightest scalar tetraquark field naturally emerges. At
nonzero density one has, beyond the usual chiral condensate, also a tetraquark
condensate. The behavior of both condensates and the restoration of chiral
symmetry at high density are studied. It is shown that this additional scalar
degree of freedom affects non negligibly the properties of the medium.

\end{abstract}

\maketitle

\section{Introduction}

The properties of strongly interacting matter at finite baryon density have
been widely investigated in the past by use of chiral models
\cite{compactstars, Mishustin:1993ub}. This paper aims to study the effect of
a light tetraquark field on such system. This subject is interesting for two
reasons: (i) many works on light mesons spectroscopy show that the light
scalar resonances below 1 GeV listed in the PDG \cite{Amsler:2008zzb} can be
successfully explained as a nonet of tetraquark states, see Refs.
\cite{Jaffe:1976ig,Maiani:2004uc,Giacosa:2006rg,Giacosa:2006tf} and refs.
therein. (ii) The nucleon can be modeled as a bound state of a quark and a
(good) diquark \cite{alkofer}. It is then conceivable that two nucleons, in
addition to the usual quark exchange leading to a quark-antiquark meson as
intermediate state, interact via an exchange of a diquark, which leads to a
tetraquark as mediator, see Fig.~\ref{figa} for a pictorial representation.

In order to perform this study we use the linear sigma model for $N_{f}=2$
described in Ref.~\cite{Gallas:2009qp}, in which the nucleon $N$ and its
chiral partner $N^{\ast}$ form a baryon doublet, $(N,N^{\ast})$, where
$N^{\ast}$ is usually identified with the resonance $N(1535)$
\cite{Amsler:2008zzb}. The doublet is introduced in the so-called mirror
assignment, first discussed in Ref.~\cite{Lee} and extensively analyzed in
Refs.~\cite{DeTar:1988kn,Jido:2001nt,Gallas:2009qp}. The particularity of the
mirror assignment is the possibility to introduce a chirally invariant mass
term $\sim m_{0}$, which does $\emph{not}$ originate from the quark
condensate. In the framework of dilatation invariant interactions, this term
originates from the condensation of two further scalar-isoscalar states: the
dilaton/glueball field and the tetraquark field. In the present work we
neglect the effect of the glueball since, due to its relatively high mass of
about $1.5$ GeV \cite{Chen:2005mg}, its exchange between nucleons is
negligible in a first approximation. On the contrary, a light tetraquark state
with a mass of about $600$ MeV and identified predominantly with the resonance
$f_{0}(600)$ is potentially very interesting for the properties of nuclear
matter. In addition, the usual (pseudo)scalar and (axial-)vector
quark-antiquark mesons are present in the model.

Applications of the parity doublet model to nuclear matter and neutron stars
were studied in
Refs.~\cite{Hatsuda:1988mv,Zschiesche:2006zj,Dexheimer:2007tn,Sasaki:2010bp}
were it was shown that, at variance with the normal linear sigma model, it is
able to describe the saturation of nuclear matter and it predicts a maximum
mass for neutron stars compatible with observations. Some problems however
arise from these studies: the nuclear matter compressibility turns out to be
larger than the measured one and the value of $m_{0}$ needed to describe
saturation is large, $\sim800$ MeV, and thus in disagreement with the results
of the analysis of Ref.~\cite{Gallas:2009qp} in which a fit of vacuum
properties resulted in $m_{0}\sim500$ MeV. Moreover the mass of the
scalar-isoscalar meson responsible for the nucleon-nucleon attraction turns
out to be very small, $m_{\sigma}\sim350$ MeV, a value that does not
correspond to any particle in the PDG.

In the present study, we remove these inconsistencies of the parity doublet
model between vacuum physics and finite density physics by regarding the
resonance $f_{0}(600)$ as a predominantly tetraquark state and the resonance
$f_{0}(1370)$ as the chiral partner of the pion, hence a predominantly
quark-antiquark state. We show that nuclear matter saturation can be correctly
reproduced with values for the compressibility compatible with data. The value
of $m_{0}$ needed for fitting high density properties is the same of the one
indicated by Ref.~\cite{Gallas:2009qp} in vacuum decays studies i.e.
$m_{0}\sim500$ MeV. Quite remarkably, the need of having two scalars within
the parity doublet model is in agreement with the implementation, in
nucleon-nucleon potentials as the one of Bonn \cite{Machleidt:2000ge}, of two
scalar mesons having masses of $\sim500$ and $\sim1200$ MeV.

The paper is organized as follows: in Sec. II we introduce the Lagrangian of
the model, in Sec. III we obtain the corresponding thermodynamic potential in
the mean field approximation, in Sec.~IV we presents our results for nuclear
matter and, finally, in Sec. V we draw our conclusions.

Our units are $\hbar=c=1$, the metric tensor is $g^{\mu\nu}=\mathrm{diag}%
(+,-,-,-)$.

\section{The parity doublet model in the baryon sector}

We present here the chirally symmetric linear sigma model with scalar,
pseudoscalar, vector, axial-vector mesons, the nucleon and its chiral partner
\cite{Gallas:2009qp}. The scalar and pseudoscalar fields are included in the
matrix
\begin{equation}
\Phi=\sum_{a=0}^{3}\phi_{a}t_{a}=(\sigma+i\eta_{N})\,t_{0}+(\vec{a}_{0}%
+i\vec{\pi})\cdot\vec{t}\;, \label{scalars}%
\end{equation}
where $\vec{t}=\vec{\tau}/2,$ with the vector of Pauli matrices $\vec{\tau}$,
and $t^{0}=\mathbf{1}_{2}/2$. Under the global $U(2)_{R}\times U(2)_{L}$
chiral symmetry, $\Phi$ transforms as $\Phi\rightarrow U_{L}\Phi
U_{R}^{\dagger},$ where $U_{L}$ and $U_{R}$ are $2\times2$ unitary matrices.
The vector and axial-vector fields are represented by the matrices
\begin{equation}
V^{\mu}=\sum_{a=0}^{3}V_{a}^{\mu}t_{a}=\omega^{\mu}\,t^{0}+\vec{\rho}^{\mu
}\cdot\vec{t},\;\,A^{\mu}=\sum_{a=0}^{3}A_{a}^{\mu}t_{a}=f_{1}^{\mu}%
\,t_{0}+\vec{a_{1}}^{\mu}\cdot\vec{t}\text{ .} \label{vectors}%
\end{equation}
From these fields we define right- and left-handed vector fields $R^{\mu
}\equiv V^{\mu}-A^{\mu}$, $L^{\mu}\equiv V^{\mu}+A^{\mu}$. Under global
$U(2)_{R}\times U(2)_{L}$ transformations, these fields transform as $R^{\mu
}\rightarrow U_{R}R^{\mu}U_{R}^{\dagger}\,,\;L^{\mu}\rightarrow U_{L}L^{\mu
}U_{L}^{\dagger}$.

The identification of mesons with particles listed in the
PDG~\cite{Amsler:2008zzb} is as follows: the fields $\vec{\pi}$ and $\eta_{N}$
correspond to the pion and the $SU(2)$ counterpart of the $\eta$ meson,
$\eta_{N}\equiv(\overline{u}u+\overline{d}d)/\sqrt{2}$, with a mass of about
$700$ MeV. This value can be obtained by \textquotedblleft
unmixing\textquotedblright\ the physical $\eta$ and $\eta^{\prime}$ mesons,
which also contain $\overline{s}s$ contributions. The vector fields
$\omega^{\mu}$ and $\vec{\rho}^{\mu}$ represent the resonances $\omega(782)$
and $\rho(770)$ and the axial-vector fields $f_{1}^{\mu}$ and $\vec{a_{1}%
}^{\mu}$ represent the resonances $f_{1}(1285)$ and $a_{1}(1260)$.
\begin{figure}[ptb]
\begin{centering}
\epsfig{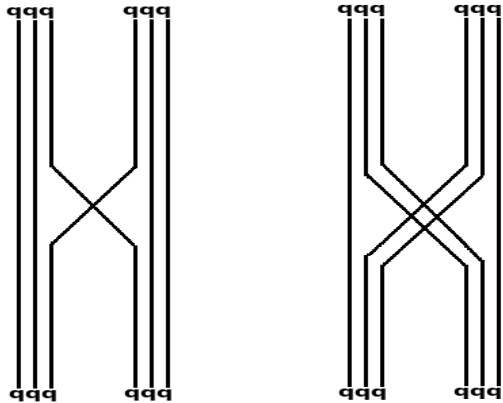}
\caption{The left picture represents the exchange of quarks between nucleons leading the
nucleon-nucleon interactions mediated by quark-antiquark mesons. The right picture
represents the exchange of diquarks leading to an additional contribution to the nucleon-nucleon potential
mediated by tetraquark mesonic states.
\label{figa} }
\end{centering}
\end{figure}Two possibilities for the identification of the $\sigma$ and
$\vec{a}_{0}$ fields exist: $\{f_{0}(600),a_{0}(980)\}$ and $\{f_{0}%
(1370),a_{0}(1450)\}$. The first assignment is however unfavoured
\cite{Parganlija:2010fz,Janowski:2011gt} (for a general discussion of the
issue of scalar mesons see also Refs. \cite{amslerrev} and refs. therein),
while the second is in agreement with the phenomenology.

The Lagrangian describing the meson fields is presented in detail in Ref.
\cite{Parganlija:2010fz,Gallas:2009qp}. For our purposes we notice that: (i)
The chiral condensate $\sigma_{0}=\left\langle 0\left\vert \sigma\right\vert
0\right\rangle =Zf_{\pi}$ emerges upon spontaneous chiral symmetry breaking in
the mesonic sector. The parameter $f_{\pi}=92.4$ MeV is the pion decay
constant and $Z$ is the wave-function renormalization constant of the
pseudoscalar fields \cite{Parganlija:2010fz,Struber:2007bm} and takes the
value $Z=1.67\pm0.2,$ which can be fixed from the process $a_{1}\rightarrow
\pi\gamma.$ (ii) In the case $N_{f}=2$ only one tetraquark state $\chi$
exists. It can be coupled to the model following
Refs.~\cite{Giacosa:2006tf,Heinz:2008cv}. (iii) The dilaton/glueball field $G$
can also be easily added to the meson sector according to the requirement of
dilatation invariance and the corresponding trace anomaly
\cite{Janowski:2011gt}. (For a general discussion of the glueball see Refs.
\cite{amslerrev,varieglue} and refs. therein.)

We now turn to the baryon sector, in which we have the baryon doublets
$\Psi_{1}$ and $\Psi_{2}$, where $\Psi_{1}$ has positive parity and $\Psi_{2}$
negative parity. In the mirror assignment they transform as follows:%

\begin{equation}
\Psi_{1R}\longrightarrow U_{R}\Psi_{1R}\,,\;\Psi_{1L}\longrightarrow U_{L}%
\Psi_{1L}\,,\;\Psi_{2R}\longrightarrow U_{L}\Psi_{2R}\,,\;\Psi_{2L}%
\longrightarrow U_{R}\Psi_{2L}\;\text{,} \label{mirror}%
\end{equation}
i.e., $\Psi_{2}$ transforms in a \textquotedblleft mirror
way\textquotedblright\ under chiral transformations \cite{Lee,DeTar:1988kn}.
These field transformations allow to write down a chiral baryonic Lagrangian:
\begin{align}
\mathcal{L}_{\mathrm{bar}}  &  =\overline{\Psi}_{1L}i\gamma_{\mu}D_{1L}^{\mu
}\Psi_{1L}+\overline{\Psi}_{1R}i\gamma_{\mu}D_{1R}^{\mu}\Psi_{1R}%
+\overline{\Psi}_{2L}i\gamma_{\mu}D_{2R}^{\mu}\Psi_{2L}+\overline{\Psi}%
_{2R}i\gamma_{\mu}D_{2L}^{\mu}\Psi_{2R}\nonumber\\
&  -\widehat{g}_{1}\left(  \overline{\Psi}_{1L}\Phi\Psi_{1R}+\overline{\Psi
}_{1R}\Phi^{\dagger}\Psi_{1L}\right)  -\widehat{g}_{2}\left(  \overline{\Psi
}_{2L}\Phi^{\dagger}\Psi_{2R}+\overline{\Psi}_{2R}\Phi\Psi_{2L}\right)
\nonumber\\
&  -(a\chi+bG)(\overline{\Psi}_{1L}\Psi_{2R}-\overline{\Psi}_{1R}\Psi
_{2L}-\overline{\Psi}_{2L}\Psi_{1R}+\overline{\Psi}_{2R}\Psi_{1L})\;\text{,}
\label{lag}%
\end{align}
where $D_{1R}^{\mu}=\partial^{\mu}-ic_{1}R^{\mu}$, $D_{1L}^{\mu}=\partial
^{\mu}-ic_{1}L^{\mu}$, and $D_{2R}^{\mu}=\partial^{\mu}-ic_{2}R^{\mu}$,
$D_{2L}^{\mu}=\partial^{\mu}-ic_{2}L^{\mu}$ are the covariant derivatives for
the nucleonic fields, with the coupling constants $c_{1}$ and $c_{2}$. The
interactions of the baryonic fields with the scalar and pseudoscalar mesons
are parametrized by $\widehat{g}_{1}$ and $\widehat{g}_{2}$. The last term in
Eq. (\ref{lag}) generates a mass term when the tetraquark field $\chi$ and the
glueball field $G$ condense:%
\begin{equation}
m_{0}=a\chi_{vac}+bG_{vac}\text{ .}%
\end{equation}
In Ref.~\cite{Gallas:2009qp} the quantity $m_{0}$ has been obtained through a
fit procedure to known experimental and lattice quantities, obtaining:%
\begin{equation}
m_{0}=460\pm130\text{ MeV .} \label{m0}%
\end{equation}
\begin{figure}[ptb]
\begin{centering}
\epsfig{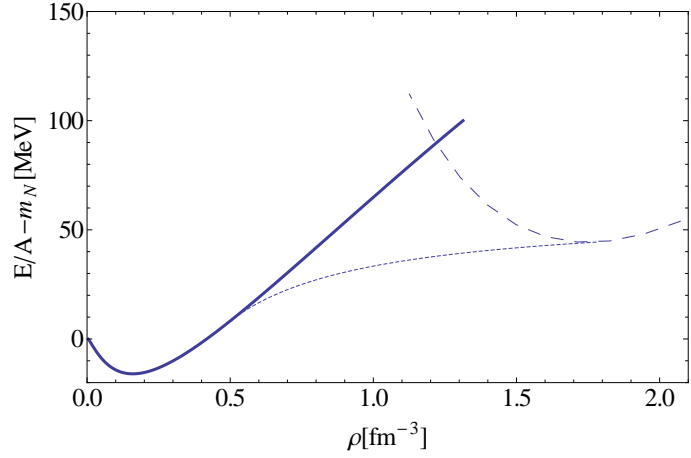}
\caption{Energy per baryon as a function of the density for the case $m_0=500$ MeV for the chiral symmetry broken solution (solid line) and the chiral
symmetry restored solution (dashed line). The dotted line represent the Maxwell construction which matches the two solutions.
The model can describe the saturation of nuclear matter.
\label{fig1} }
\end{centering}
\end{figure}In this paper we work under the simplified assumption $b=0$, i.e.
the parameter $m_{0}$ is saturated by the tetraquark condensate and the
corresponding nucleon-nucleon interaction is depicted in Fig.~\ref{figa},
right side. In this context the coupling constant $a$ is fixed as soon as the
tetraquark condensate is specified, see next section. \begin{figure}[ptb]
\begin{centering}
\epsfig{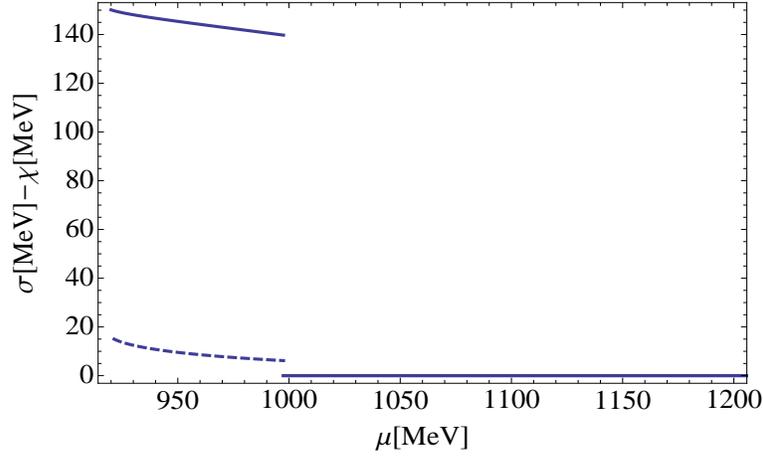}
\caption{Mean $\sigma$ (solid line) and $\chi$ (dashed line) fields. A strong first order phase transition
is present at $\mu \sim 1$ GeV. The transition point is the same for the two fields. Note, the minimal value for the chemical potential
corresponds to nuclear matter.
\label{fig2} }
\end{centering}
\end{figure}The physical fields $N$ and $N^{\ast}$ are related to the spinors
$\Psi_{1}$ and $\Psi_{2}$ through:
\begin{equation}
\Psi_{1}=\frac{1}{\sqrt{2\cosh\delta}}(Ne^{\delta/2}+\gamma_{5}N^{\ast
}e^{-\delta/2})\;,\text{ }\Psi_{2}=\frac{1}{\sqrt{2\cosh\delta}}(\gamma
_{5}Ne^{-\delta/2}-N^{\ast}e^{\delta/2})\;. \label{rotationfelder}%
\end{equation}
The masses of the nucleons are obtained by diagonalizing the corresponding
mass matrix in the Lagrangian. As function of the two condensates
$\sigma_{vac}$ and $\chi_{vac}$ they read:
\begin{equation}
m_{N,N^{\ast}}=\sqrt{\left(  \frac{\widehat{g}_{1}+\widehat{g}_{2}}{4}\right)
^{2}\sigma_{vac}^{2}+(a\chi_{vac})^{2}}\pm\frac{\widehat{g}_{1}-\widehat
{g}_{2}}{4}\sigma_{vac}.
\end{equation}

\section{Mean field approximation}

We now turn to the effective model in the mean field approximation suitable
for describing nuclear matter. We thus retain only those fields which are
relevant for such a study: the scalar fields $\sigma$ and $\chi$ and the
vector field $\omega$. The effective Lagrangian is given by:%

\begin{align}
&  \mathcal{L}_{eff}=\frac{1}{2}\partial_{\mu}\sigma\partial^{\mu}\sigma
+\frac{1}{2}\partial_{\mu}\chi\partial^{\mu}\chi-\frac{1}{4}(\partial_{\mu
}\omega_{\nu}-\partial_{\nu}\omega_{\mu})^{2}\nonumber\\
&  +\frac{1}{2}m^{2}\sigma^{2}+\frac{1}{2}m_{1}^{2}\omega_{\mu}\omega^{\mu
}-\frac{1}{2}m_{\chi}^{2}\chi^{2}-\frac{\lambda}{4}\sigma^{4}+g\chi\sigma
^{2}+\varepsilon\sigma\nonumber\\
&  +\overline{\Psi}_{1}i\gamma_{\mu}\partial^{\mu}\Psi_{1}+\overline{\Psi}%
_{2}i\gamma_{\mu}\partial^{\mu}\Psi_{2}-\frac{\widehat{g}_{1}}{2}%
\overline{\Psi}_{1}\sigma\Psi_{1}-\frac{\widehat{g}_{2}}{2}\overline{\Psi}%
_{2}\sigma\Psi_{2}\nonumber\\
&  -g_{\omega}^{(1)}\overline{\Psi}_{1}i\gamma_{\mu}\omega^{\mu}\Psi
_{1}-g_{\omega}^{(2)}\overline{\Psi}_{2}i\gamma_{\mu}\omega^{\mu}\Psi
_{2}+a\chi(\overline{\Psi}_{2}\gamma_{5}\Psi_{1}-\overline{\Psi}_{1}\gamma
_{5}\Psi_{2})\text{ .} \label{lageff}%
\end{align}
If we take their vacuum expectation values, $\sigma\rightarrow\sigma
+\sigma_{vac}$ and $\chi\rightarrow\chi+\chi_{vac}$, the potential for the
fields $\sigma$ and $\chi$ reads:%

\[
V(\sigma,\chi)=\frac{1}{2}%
\begin{pmatrix}
\chi & \sigma
\end{pmatrix}%
\begin{pmatrix}
m_{\chi}^{2} & -2g\sigma_{vac}\\
-2g\sigma_{vac} & m_{\sigma}^{2}%
\end{pmatrix}%
\begin{pmatrix}
\chi\\
\sigma
\end{pmatrix}
\text{ }.
\]
A non diagonal term, which is proportional to the parameter $g$ and mixes the
quarkonium field $\sigma$ and the tetraquark field $\chi,$ has emerged
\cite{footnote}. The physical fields, denoted as $h$ and $s,$ are obtained
through a standard diagonalization:
\[%
\begin{pmatrix}
h\\
s
\end{pmatrix}
=%
\begin{pmatrix}
\mathrm{cos}\theta_{0} & \mathrm{sin}\theta_{0}\\
-\mathrm{sin}\theta_{0} & \mathrm{cos}\theta_{0}%
\end{pmatrix}%
\begin{pmatrix}
\chi\\
\sigma
\end{pmatrix}
\text{ },\text{ }\theta_{0}=\frac{1}{2}\mathrm{arctan}\frac{4g\sigma_{vac}%
}{m_{\sigma}^{2}-m_{\chi}^{2}}\text{ .}%
\]
Then the physical masses of the scalar states are:
\begin{align}
&  m_{h}^{2}=m_{\chi}^{2}\mathrm{cos}^{2}\theta_{0}+m_{\sigma}^{2}%
\mathrm{sin}^{2}\theta_{0}-2g\sigma_{vac}\mathrm{sin}(2\theta_{0}),\nonumber\\
&  m_{s}^{2}=m_{\sigma}^{2}\mathrm{cos}^{2}\theta_{0}+m_{\chi}^{2}%
\mathrm{sin}^{2}\theta_{0}+2g\sigma_{vac}\mathrm{sin}(2\theta_{0}).
\end{align}
We identify here the predominately tetraquark state $h$ with $f_{0}(600)$ (the
vacuum mass $m_{h}=600$ MeV is used) and the predominately quarkonium state
$s$ with $f_{0}(1370)$ (the vacuum mass $m_{s}=1300$ MeV is used) .
\begin{figure}[ptb]
\begin{centering}
\epsfig{file=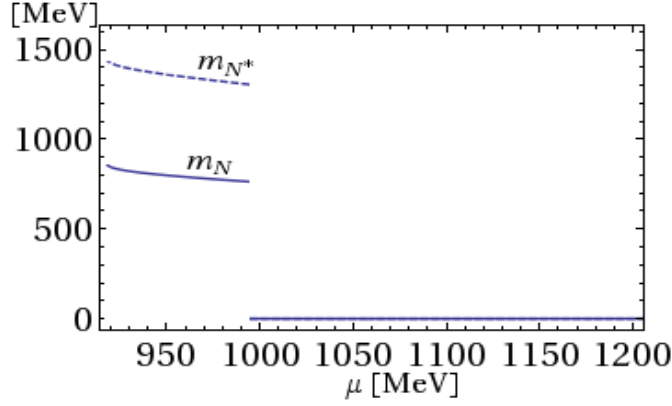,height=6cm,width=10cm}
\caption{Masses of the nucleons (solid line) and their chiral partners (dashed line) as function of the chemical potential.
\label{fig3} }
\end{centering}
\end{figure}
We now turn to the thermodynamic potential in the mean field approximation:%

\[
\frac{\Omega}{V}=-\mathcal{L_{\mathrm{M}}}+\sum_{i}\frac{\gamma_{i}}%
{(2\pi)^{3}}\int_{0}^{p_{F_{i}}}\,d^{3}p\,(E_{i}^{\ast}(p)-\mu_{i}^{\ast})\ ,
\]
where the mesons term reads
\begin{equation}
\mathcal{L_{\mathrm{M}}}=\frac{1}{2}m_{1}^{2}\omega_{0}^{2}+\frac{1}{2}%
m\,^{2}\sigma^{2}-\frac{\lambda}{4}\sigma^{4}+\epsilon\sigma+g\chi\sigma
^{2}-\frac{1}{2}m_{\chi}^{2}\chi^{2}.\nonumber
\end{equation}
The index $i={N,N^{\ast}}$ denotes the nucleon type (positive and negative
parity nucleons), $\gamma_{i}=2\times2$ is the fermionic degeneracy (spin and
isopsin), $p_{F_{i}}$ are the Fermi momenta, $E_{i}^{\ast}(p)=\sqrt{p_{i}%
^{2}+{m_{i}}^{2}}$, and $\mu_{i}^{\ast}=\mu_{i}-g_{\omega}\omega_{0}%
=\sqrt{p_{F_{i}}^{2}+{m_{i}}^{2}}$. The single particle energy of each parity
partner $i$ is given by $E_{i}(p)=E_{i}^{\ast}(p)+g_{\omega}\omega_{0}$. The
mean mesons fields are obtained by minimizing the thermodynamic potential,
i.e. by solving the following system of non-linear equations:

\begin{align}
&  \frac{\partial(\Omega/V)}{\partial\sigma}\;=\;\lambda\sigma^{3}-m^{2}%
\sigma-\epsilon-2g\chi\sigma+\rho_{N}^{\ast}(\sigma,\omega_{0},\chi
)\frac{\partial m_{N}}{\partial\sigma}+\rho_{N^{\ast}}^{\ast}(\sigma
,\omega_{0},\chi)\frac{\partial m_{N}^{\ast}}{\partial\sigma}=0\text{
}\nonumber\\
&  \frac{\partial(\Omega/V)}{\partial\omega}=m_{\omega}^{2}\omega
_{0}-g_{\omega,N}\rho_{N}(\sigma,\omega_{0},\chi)-g_{\omega,N^{\ast}}%
\rho_{N^{\ast}}(\sigma,\omega_{0},\chi)=0\text{{} ,}\nonumber\\
&  \frac{\partial(\Omega/V)}{\partial\chi}=-g\sigma^{2}+m_{\chi}^{2}\chi
+\rho_{N}^{\ast}(\sigma,\omega_{0},\chi)\frac{\partial m_{N}}{\partial\chi
}+\rho_{N^{\ast}}^{\ast}(\sigma,\omega_{0},\chi)\frac{\partial m_{N}^{\ast}%
}{\partial\chi}=0\text{ .}%
\end{align}
The scalar densities $\rho_{i}^{\ast}$ and the baryon densities $\rho_{i}$ are
given by the expressions:
\begin{equation}
\rho_{i}^{\ast}=\gamma_{i}\int_{0}^{p_{F_{i}}}\frac{d^{3}\vec{p}}{(2\pi)^{3}%
}\frac{m_{i}}{\sqrt{p_{i}^{2}+m_{i}^{2}}}=\frac{m_{i}}{\pi^{2}}\left[
p_{F_{i}}E_{F_{i}}^{\ast}-m_{i}^{2}\ln\left(  \frac{p_{F_{i}}+E_{F_{i}}^{\ast
}}{m_{i}}\right)  \right]  \text{ ,}%
\end{equation}%
\begin{equation}
\rho_{i}=\gamma_{i}\int_{0}^{p_{F_{i}}}\frac{d^{3}\vec{p}}{(2\pi)^{3}}%
=2\frac{p_{F_{i}}^{3}}{3\pi^{2}}\;.
\end{equation}
The parameters of the model are given by \cite{Parganlija:2010fz,Heinz:2008cv}%
:
\begin{equation}
\lambda=\frac{1}{2(Zf_{\pi})^{2}}\left(  m_{\sigma}^{2}-\frac{m_{\pi}^{2}%
}{Z^{2}}\right)  +\frac{2g^{2}}{m_{\chi}^{2}}\;,\text{ }m^{2}=\frac{1}%
{2}\left(  m_{\sigma}^{2}-3\frac{m_{\pi}^{2}}{Z^{2}}\right)  \;,\text{
}\varepsilon=\frac{f_{\pi}m_{\pi}^{2}}{Z},\,\,a=\frac{m_{0}m_{\chi}^{2}%
}{g(Zf_{\pi})^{2}}\text{ .}%
\end{equation}
Numerically one has $m_{\pi}=139$ MeV and $f_{\pi}=92.4$ MeV. The parameter
$Z$ is the wave function renormalization of the pseudoscalar fields
\cite{Parganlija:2010fz,Struber:2007bm}. It is the only `remnant' of the
(axial-)vector mesons of the model. However, it should be stressed that the
value of $Z=1.67$ strongly affects the properties of nuclear matter, for
instance reducing the value of the chiral symmetry explicit breaking term, as
we will show in the following. As already discussed in
Refs.~\cite{Parganlija:2010fz}, the presence of (axial-)vector degrees of
freedom in a chiral framework has non-negligible consequences on the
(pseudo)scalar sector as well.
\begin{figure}[ptb]
\begin{centering}
\epsfig{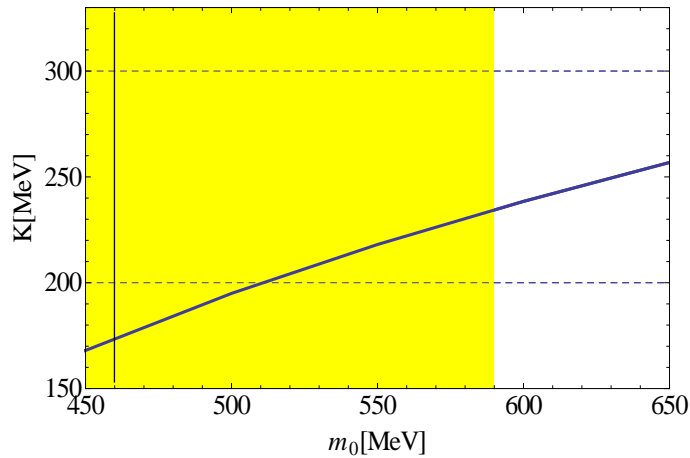}
\caption{Compressibility as a function of $m_0$ (solid line) and allowed range for $K$ as obtained by phenomenology (dashed lines). The constraint
on the compressibility, together with vacuum decays properties, fixes $m_0$ to be in the range $500$-$600$ MeV. The result ($460\pm130$) MeV is shown
(gray (yellow online)  band).
\label{fig4} }
\end{centering}\end{figure}%
The coupling constants of the nucleons and the scalar $\sigma$ meson are
uniquely determined by the vacuum properties, $m_{0}=460\pm130$ MeV,
$m_{N_{vac}}=938$ MeV and $m_{N_{vac}^{\ast}}=1535$ MeV: $\widehat{g}%
_{1}=11.0\pm1.5,\;\widehat{g}_{2}=18.8\pm2.4$ \cite{Gallas:2009qp}. Finally,
$g$ parametrizes the interactions of the tetraquark field with pions and the
parameters $g_{\omega}^{(1)},$ $g_{\omega}^{(2)}$ describe the interactions of
the $\omega$ meson with the nucleons. For sake of simplicity we assume that
$g_{\omega}^{(1)}=g_{\omega}^{(2)}=g_{\omega}$ and thus, $g_{\omega
,N}=g_{\omega,N^{\ast}}$. The quantities $g$ and $g_{\omega}$ are obtained
from the following constraints of nuclear matter at saturation:
\begin{equation}
\frac{\partial}{\partial p_{F_{N}}}\left(  \frac{E}{A}-m_{N_{vac}}\right)
\biggr|_{p_{F_{N}}=p_{F_{0}}}=0\;\;\;\text{and}\;\;\left(  \;\frac{E}%
{A}-m_{N_{vac}}\right)  \biggr|_{p_{F_{N}}=p_{F_{0}}}=-16\text{ }%
\mathrm{MeV}\,,
\end{equation}
where $E/A$ is the energy per nucleon and $p_{F_{N}}$ is the Fermi momentum of
$N$. At the saturation point the latter equals value $p_{F_{N}}=p_{F_{0}}=258$ MeV.

\section{Results}

Let us fix $m_{0}$ to the intermediate value $m_{0}=500$ MeV. From Eqs.~(10)
and (15) we can determine the four free parameters of the model which turn out
to be: $g_{\omega}=4.87$, $g=450$ MeV, $m_{\sigma}=1294$ MeV, $m_{\chi}=612$
MeV (the latter two quantities generate the previously mentioned massed
$m_{h}=600$ MeV and $m_{s}=1300$ MeV). Quite remarkably the values of the
masses of the two scalar mesons of our model are very close the values
obtained within the \textquotedblleft Bonn parametrization\textquotedblright%
\ of the nucleon-nucleon potential extracted by scattering data
\cite{Machleidt:2000ge}. In Fig.~\ref{fig1}, we show the energy per baryon
$E/A$ as a function of the baryon density $\rho$: one can notice the typical
feature of the saturation of nuclear matter as a minimum of $E/A$ located at
$\rho=\rho_{0}=0.16$ fm$^{-3}$ and with a nucleon binding energy of $16$ MeV.

The value of the compressibility, computed through the derivative of the
pressure $P$ with respect to the density:
\begin{equation}
K=9\frac{\partial P}{\partial\rho}\biggr|_{\rho=\rho_{0}}=194\text{ MeV ,}%
\end{equation}
a value which is very close to the standard range indicated by phenomenology
$K=200$-$300$ MeV \cite{Youngblood:2004fe,Hartnack:2007zz}. In the figure we
also display the energy per baryon for the chiral symmetry restored solution
of the mean field equations (dashed line) and the interpolating Maxwell
construction (dotted line) which we will explain in the following.

At high densities, our model predicts a strong first order chiral phase
transition occurring at a chemical potential $\mu\sim1$ GeV as displayed in
Fig.~\ref{fig2} where we show the mean scalar mesons fields ($\sigma$ with
solid line and $\chi$ with dashed line). The big jump in both the $\sigma$ and
$\chi$ fields is computed by using a Maxwell construction: at a fixed value of
the chemical potential the mean mesons fields are the solutions of Eqs.~(11)
which minimize the thermodynamic potential. In the energy per baryon - density
diagram, the Maxwell construction corresponds to an hyperbolic branch (see
Fig.~2, dotted line) given by the thermodynamic equation relating the pressure
and the energy density: $e/\rho=\mu_{crit}-p_{crit}/\rho$ where $\mu_{crit}$
and $p_{crit}$ are the critical chemical potential and pressure.
\begin{figure}[ptb]
\begin{centering}
\epsfig{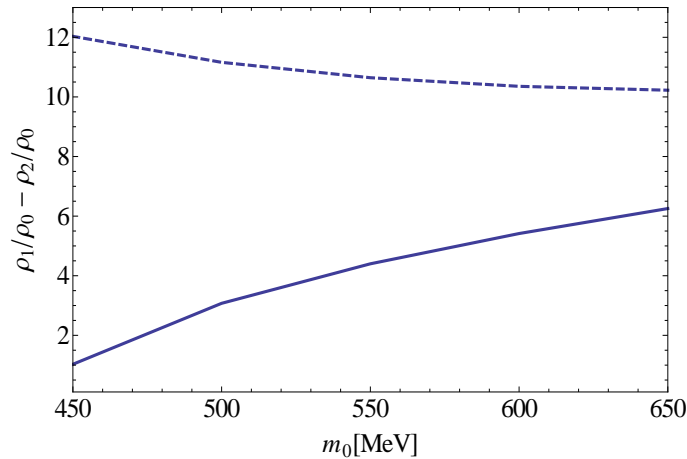}
\caption{Critical densities associated with the chiral phase transition as functions of $m_0$. $\rho_1$ (solid line) corresponds
to the onset of the phase transition and $\rho_2$ (dashed line) to the end (the densities are referred to the saturation density $\rho_0$).
The first order phase transition becomes weaker for large values of $m_0$.
\label{fig5} }
\end{centering}
\end{figure}
Notice that in absence of the tetraquark field and without the corrections
related to $Z$ \cite{Zschiesche:2006zj}, the phase transition, while being
still a first order, is much weaker. The parameter $Z=1.67$ indeed suppresses
the chiral symmetry explicit breaking term and enhances the vacuum expectation
value of $\sigma$, thus making the results for the chiral phase transition
similar to the ones obtained in the chiral limit. In our model the chiral
phase transition is accompanied by the appearance of the chiral partners of
the nucleon. While at low density only the nucleons are present, after the
phase transitions both, the nucleons and their chiral partners, contribute the
same amount. As one can notice in Fig.~\ref{fig3}, in which we show the masses
of the nucleons and their chiral partners as a function of the chemical
potential, the nucleons become massless after the phase transition. This result
is different from the one of Ref.~\cite{Zschiesche:2006zj} in which the nucleon mass,
at high density, saturates to the fixed value of $m_0$. On the other hand, in our model, $m_0$ itself 
is a function of the density and at the chiral phase transtion it vanishes
since the tetraquark condensate vanishes (see Eq.~5).

As a further step in our study we investigate the effect of varying the
parameter $m_{0}$. For each value of $m_{0}$ we compute the values of the four
parameters described before: while we can describe saturation for the whole
range of Eq.~(\ref{m0}), the value of $K$ turns out to be very small for small
values of $m_{0}$. We show the corresponding result in Fig.~\ref{fig4}
together with the range of $K$ indicated by phenomenology (dashed lines). By
combing the information of Eq.~(6) and the constraint obtained from the value
of $K$ we can further restrict the range of the allowed values for the
\textquotedblleft bare mass\textquotedblright\ of the nucleon $500\lesssim
m_{0}\lesssim600$ MeV.

Finally, in Fig.~\ref{fig5}, we show the values of the critical densities
$\rho_{1}$ and $\rho_{2}$ which correspond to the onset and the end of the
chiral phase transition. One can notice that the strength of the phase
transition decreases for large values of $m_{0}$; within the allowed range of
$m_{0}$ the density jump at the phase transition is anyway large : $\rho
_{2}/\rho_{1}=2$-$4$. Remarkably, recent lattice results \cite{li} show for a
temperature $T\sim0.8T_{c}$ a rather broad range of density for the mixed
phase, $\rho_{2}/\rho_{1}\sim2.$ Such studies are presently limited for
temperatures $T\gtrsim0.8T_{c},$ therefore a quantitative comparison with our
results is premature. Nevertheless, the lattice simulations seem to indicate
that the interval for the mixed increase further when diminishing the
temperature $T.$ Future detailed analysis on this issue is surely interesting.

\section{Conclusions}

In this work we have used the chiral model developed in Ref.
\cite{Gallas:2009qp} to study nuclear matter properties. The model contains,
in addition to the standard quark-antiquark fields with (pseudo)scalar and
(axial-)vector quantum numbers, also a light scalar-isoscalar tetraquark
field. In the framework of the mirror assignment the latter can be easily
coupled to the nucleon by using the general requirement of dilatation
invariant interactions and contributes to the nucleon-nucleon interaction, see Fig.~1.

Similarly to the results of Ref.~\cite{Heinz:2008cv} at finite temperature and
vanishing density, we find that also at nonzero density a light tetraquark
field has a strong influence on the medium properties of the system due to the
interplay of two condensates, the tetraquark and the chiral (quark-antiquark)
condensates. Interestingly, the described scenario and also the values of the
coupling constants are in agreement with the work of
Ref.~\cite{Machleidt:2000ge}, where two scalar fields are needed to describe
nucleon-nucleon scattering data. As described in Ref. \cite{Bonanno:2011yr},
in such a scenario nuclear matter is a prerogative of our world with three
colors ($N_{c}=3),$ but would cease to exist as soon as a larger number of
colors is considered.

An important parameter of the model is $m_{0}$, which describes the
contribution to the nucleon mass which does not stem from the chiral
condensate and which, in the present study, is saturated by the tetraquark
condensate. Its value has been fixed in the vacuum as $m_{0}=460\pm130$ MeV.
It is then remarkable that the compressibility is in agreement with the
experiment in a compatible range of $m_{0}$ (see Fig.~5), thus showing that
vacuum results and nonzero density properties can be understood within the
same theoretical model.

Further studies along the direction of the present work can be performed: (i)
the contribution of the gluon condensate to $m_{0}$ and of the
dilaton/glueball field to the nucleon-nucleon interaction should be included.
Although the glueball is heavier ($\sim1.5$ GeV) and should not affect the
interaction of two nucleons, its inclusion is important in a theoretical
framework which is based on the dilatation invariance of QCD and on its
anomalous breaking. (ii) The study of asymmetric nuclear matter and its
application to neutron stars is also important since, recently, very massive
neutron stars have been discovered which can give useful constraints for the
stiffness of the equation of state. Moreover, also the symmetry energy of
nuclear matter at high density is a crucial quantity that we can investigate
by introducing in our model the scalar isovector $a_{0}(980)$, as done in
Ref.~\cite{Liu:2001iz}, which in our assignment is also a tetraquark. (iii)
The extension of the model to $N_{f}=3$ is currently under investigation. The
effects of nonzero density and temperature of the complete model with all the
relevant degrees of freedoms represents an interesting outlook of the present work.

\bigskip The work of G.P. is supported by the Deutsche Forschungsgemeinschaft
(DFG) under Grant No. PA 1780/2-1. We thank D.H. Rischke, J. Schaffner-Bielich
and L. Bonanno for valuable discussions.

\bibliographystyle{plain}
\bibliography{references}

\begin{thebibliography}{99}
\bibitem {footnote}The term responsible for the tetraquark-quarkonium
interaction, and therefore also for the mixing, is $g\chi\sigma^{2}$ in Eq.
(\ref{lageff}). It was introduced in Ref. \cite{Giacosa:2006tf} because it is
dominant in the large-$N_{c}$ expansion. (Terms of the kind $\chi^{2}%
\sigma^{2}$ would also possible, but they are here neglected because
large-$N_{c}$ suppressed.) Note also that the term $g\chi\sigma^{2}$ is the
only non-dilatation invariant term in the chiral limit. In a fully dilatation
invariant context one should write down the term $\alpha G\chi\sigma^{2}$
where $G$ is the dilaton field. Upon condensation of the latter one obtains
$g=\alpha G_{0}.$ As long as the fluctuations of the dilaton are neglected it
is justified to work with $g\chi\sigma^{2}.$
\end{thebibliography}

\end{document}